\documentclass[sigconf,bookmarks=false,authorversion,nonacm]{acmart}
\usepackage[latin9]{inputenc}
\usepackage{algorithm2e}
\usepackage{amstext}
\usepackage{amsthm}
\usepackage{graphicx}
\usepackage{tikz}

\makeatletter

\providecommand{\tabularnewline}{\\}

\setcitestyle{super,sort&compress}
\newcommand\copyrighttext{This paper has been published in the Proceedings of the 23rd International ACM Conference on Modeling, Analysis and Simulation of Wireless and Mobile Systems (MsWiM 2020), Alicante, Spain, November 16--20, 2020. Version of record DOI: 10.1145/3416010.3423238 .}
\newcommand\copyrightnotice{%
\begin{tikzpicture}[remember picture,overlay]
\node[anchor=south,yshift=40pt] at (current page.south) {\fbox{\parbox{\dimexpr\textwidth-\fboxsep-\fboxrule\relax}{\copyrighttext}}};
\end{tikzpicture}%
}
\authorsaddresses{}

\setcitestyle{numbers}

\makeatother

\begin{document}

\fancyhead{}

\title{Optimal Popularity-based Transmission Range Selection for
D2D-supported Content Delivery}

\author{Loreto Pescosolido}

\affiliation{\institution{Italian National Research Council, Institute
for Informatics and Telematics (CNR-IIT)}\streetaddress{Via Giuseppe
Moruzzi 1}\city{Pisa} \postcode{I-00524}\country{Italy}} 

\email{loreto.pescosolido@iit.cnr.it}

\author{Andrea Passarella}

\affiliation{\institution{Italian National Research Council, Institute
for Informatics and Telematics (CNR-IIT)}\streetaddress{Via Giuseppe
Moruzzi 1}\city{Pisa} \postcode{I-00524}\country{Italy}} 

\email{andrea.passarella@iit.cnr.it} 

\author{Marco Conti}

\affiliation{\institution{Italian National Research Council, Institute
for Informatics and Telematics (CNR-IIT)}\streetaddress{Via Giuseppe
Moruzzi 1}\city{Pisa} \postcode{I-00524}\country{Italy}} 

\email{marco.conti@iit.cnr.it}

\begin{abstract}
Considering device-to-device (D2D) wireless links as a virtual extension
of 5G (and beyond) cellular networks to deliver popular contents has
been proposed as an interesting approach to reduce energy consumption,
congestion, and bandwidth usage at the network edge. In the scenario
of multiple users in a region independently requesting some popular
content, there is a major potential for energy consumption reduction
exploiting D2D communications. In this scenario, we consider the problem
of selecting the maximum allowed transmission range (or equivalently
the maximum transmit power) for the D2D links that support the content
delivery process. We show that, for a given maximum allowed D2D energy
consumption, a considerable reduction of the cellular infrastructure
energy consumption can be achieved by selecting the maximum D2D transmission
range as a function of content class parameters such as popularity
and delay-tolerance, compared to a uniform selection across different
content classes. Specifically, we provide an analytical model that
can be used to estimate the energy consumption (for small delay tolerance)
and thus to set the optimal transmission range. We validate the model
via simulations and study the energy gain that our approach allows
to obtain. Our results show that the proposed approach to the maximum
D2D transmission range selection allows a reduction of the overall
energy consumption in the range of 30\% to 55\%,
compared to a selection of the maximum D2D transmission range oblivious
to popularity and delay tolerance.
\end{abstract}


\maketitle
\copyrightnotice

~\newpage
\section{Introduction}

Device-to-Device (D2D) communications have been considered as a promising
technology for increasing the efficiency of cellular networks since
Release 12 of LTE standard \cite{LTE_REL_12}. Over the last few years,
many studies on 5G networks and next generation networks have investigated
the integration of D2D communications in the cellular network as a
viable tool to support the delivery, to mobile users, of \emph{popular
contents}, i.e., content that are requested by a relevant percentage
of the users \cite{Asadi2014,Rebecchi2015}. The application domains
for the D2D-supported delivery of popular contents include, besides
proximity services, human driven data traffic demand (e.g., news feeds,
tweets, popular videos, etc.) as well as automated downloads triggered,
e.g., by smart cities applications running in the user equipments
(UEs).

In the related literature, D2D communications are more and more being
considered as a part of cellular network fabric, i.e., as a viable
option that can be used to satisfy a share of the traffic load, \cite{Whitbeck2012,Rebecchi2016,Sciancalepore2016,Belouanas2019},
provided that the achieved QoS is compatible with the requirements
of that traffic share. The reduction of the energy consumption of
cellular networks' base stations (BSs), along with congestion and
spectrum usage minimisation, are all major goals of the protocol designs
following this approach. Indeed, energy consumption at the BSs induces
a relevant part of the OPEX of mobile network operators, \cite{Auer2011},
and its reduction is a constant goal in the design of radio access
networks\footnote{Besides the energy expenditure required to keep the BSs
active (which does not depend on the amount of user data transmitted)
a considerable share of this expenditure is associated to the amount
of user data actually transmitted in downlink operations, including
the cost of the energy for both processing and RF transmission.}. With D2D-support in place, it is important that a high percentage
of users receive contents by neighboring devices, instead of obtaining
it from the BS. Previous studies have shown that delay tolerant traffic
demand is particularly suited to be satisfied through D2D communications.
In fact, the probability of \emph{offloading} the content delivery
process to D2D communications benefits from the additional degree
of freedom associated with the possibility, for the users, to wait
for some time before receiving a content they have requested. One
of the major factors affecting the willingness of the users to participate
in the D2D content delivery process is the energy consumed by the
devices for the D2D operation. Thus, any sensible D2D offloading strategies
pursue the objective of minimising the infrastructure-to-device (I2D)
energy consumption while keeping the devices' consumption under a
tolerable level. A similar problem could be casted in terms of fixing
a target reduction of the I2D energy consumption and attempt to achieve
it with the least possible D2D energy consumption. Finally, a global
energy minimisation could be pursued, i.e., attempting to minimise
the overall energy expenditure resulting from both I2D and D2D communications.

In the attempt to minimise energy consumption, there is an obvious
interplay between the offloading efficiency\footnote{Equivalent terms used in the literature are ``offloading probability'' and ``cache hit probability''.}, defined here as the percentage of content requests satisfied through D2D communications, and the transmit power used for them. Despite the relevant body of works covering the topic of D2D-supported cellular networks, most approaches optimise relevant parameters only at a specific layer in the protocol stack. This is the case, for example, of approaches focusing on physical or layer-2 parameters such as the transmit power, which do not consider higher level information (such as the popularity
of contents being requested), and vice versa. 

In this work, we contribute to fill this gap by showing the advantage
of optimising parameters through a cross-layer approach, mixing physical
layer parameters and application-level parameters. Specifically, we
optimise the transmit power of D2D communications (to achieve minimal
energy consumption) based on the popularity of contents request and
the probability they are requested over time after they are generated.
More precisely, we argue that the selection of the maximum D2D transmission
range should depend on traffic type features as the content popularity,
defined as the percentage of users interested in contents classified
in a certain traffic type, the delay tolerance for the considered
traffic type, and the typical request intensity profile for that traffic
type, which represents how the intensity of requests from the population
of users for a particular content evolves in time. We start presenting
a general framework for formulating the above defined energy minimisation
problems. Then we propose an analytical model for non-delay tolerant
traffic that allows to compute the optimal maximum D2D transmission
range as a function of popularity and request intensity profile of
the different traffic types. We extend our analysis to delay tolerant
traffic, for which we exploit simulation results showing the additional
margin for energy consumption reduction entailed by this type of traffic.
Finally we compare the proposed strategy of selecting the maximum
D2D range as a function of traffic parameters against a uniform maximum
distance selection, in the presence of a mixed traffic demand. This
benchmark represents the case of optimisation neglecting traffic-related
information. Our results show that, for the scenario parameters considered
in our performance evaluation, when minimising the overall (D2D +
I2D) energy consumption, the selective optimisation consumes at least
35\% less energy then a system in which a uniform maximum transmission
range is used for all content types.

The paper is organized as follows. In Section~\ref{sec:Related-work}
we position our contribution with respect to existing literature.
In Section~\ref{sec:System-model} we present our system model in
terms of network model, considered data traffic characteristics, and
D2D operation. In Section~\ref{sec:Problem-formulation} we formulate
the considered energy minimisation problem and present the proposed
analytical model. In Section~\ref{sec:Performance-evaluation} we
validate the model and evaluate the performance of the proposed techniques
in terms of energy consumption through extensive simulation results.
Section~\ref{sec:Conclusion} summarizes our results and points at
future research directions.

\section{Related work\label{sec:Related-work}
}

The literature on optimisation aspects at different layers of the
network architecture for the coexistence of D2D communications within
a cellular network, and on D2D offloading in particular, is vast.
One of the parameters that are typically subject to optimisation is
the transmit power of the devices. In layer 2-oriented studies, the
D2D transmit power control is used to enable an optimal radio resource
sharing of D2D communications with I2D (and viceversa) ones by mitigating
the interference among transmissions that share the same resources,
assuming the two types of links are eligible for using the same resources
\cite{Doppler2009,Yu2009}. In most network-level studies, the transmission
range settings are taken as an input to the problem. To the best of
our knowledge, linking the selection of the maximum D2D transmission
range to specific classes of traffic demand is an approach that has
not yet been investigated in the literature.

In the networking community, D2D offloading problems have been framed
under different assumptions on the system model. The interested reader
can check \cite{Rebecchi2015} for large spectrum classification.
Scenarios assuming synchronous request make it possible to organize
the content delivery by partially leveraging multicast, see e.g.,
\cite{Rebecchi2015b}. In this work, we consider requests arising
from users independently of each other (except for the fact that they
are all subsequent to a content generation). Pioneering studies date
back to the works \cite{Whitbeck2011,Whitbeck2012}. Our interest
is in the energy savings that can be achieved without requiring the
network to explicitly organize the diffusion of each content among
the interested users. Therefore, we do not consider selective content
injection \cite{Rebecchi2016,Sciancalepore2016}, and multicast planning
operation either. While these strategies are without question worth
for reducing the network resources usage, they require a specific
planning for each content. The solution presented in this paper requires
the network controller to execute minimal operation, as the content
diffusion process evolves accordingly to the spontaneous requests
from the users. We do require, however, that the cellular network
has a certain degree of at least \emph{monitoring} how popular contents
spread within the population of users. An assumption we make, which
is also popular in works dealing with the 5G and beyond network architecture
(see e.g. \cite{Lee2020}) is that the network has knowledge of the
nodes positions and of which contents the users are caching at any
time. This is used to identify neighboring nodes and instruct them
to hand content of mutual interest to each other.

Regarding the characterisation of the content classes, an important
feature in our work is the request intensity profile, i.e., a function
of time which tells how the number of requests (per unit time) evolves
after the generation of a content. For the scope of this work, we
target contents that become of interest for many users in a matter
of tens of minutes, according to the already mentioned application
domains. While there are many studies covering the topic of how the
popularity of a given content evolves in time, they are mostly targeted
at videos shared on popular platforms, for which popularity evolves
on timescales in the order of days, or even months \cite{Korosi2011,Garroppo2018}.
This type of works, however, are not focused on capturing the initial,
raising phase of the popularity dynamics, as they are more interested
in the long-term behavior. In \cite{Mangili2016}, the authors leverage
the rank-and shift model to establish raises and decays in relative
popularity of the contents in a content library. While their setup
is completely different from our one, this model confirms a typical
behavior of the popularity dynamics, in which there is an initially
slowly raising phases, followed by an acceleration and deceleration
in the growth of the popularity, until a peak of request is reached.
In this work, we adopt a parametrical approach, similar to that in
\cite{Avramova2009} which, however, is still focused on the tail
of the distribution of the requests. 

\section{System model\label{sec:System-model}}

\subsection{Network model and energy consumption\label{subsec:Network-model}}

\begin{figure}
\begin{centering}
\includegraphics[width=1\columnwidth]{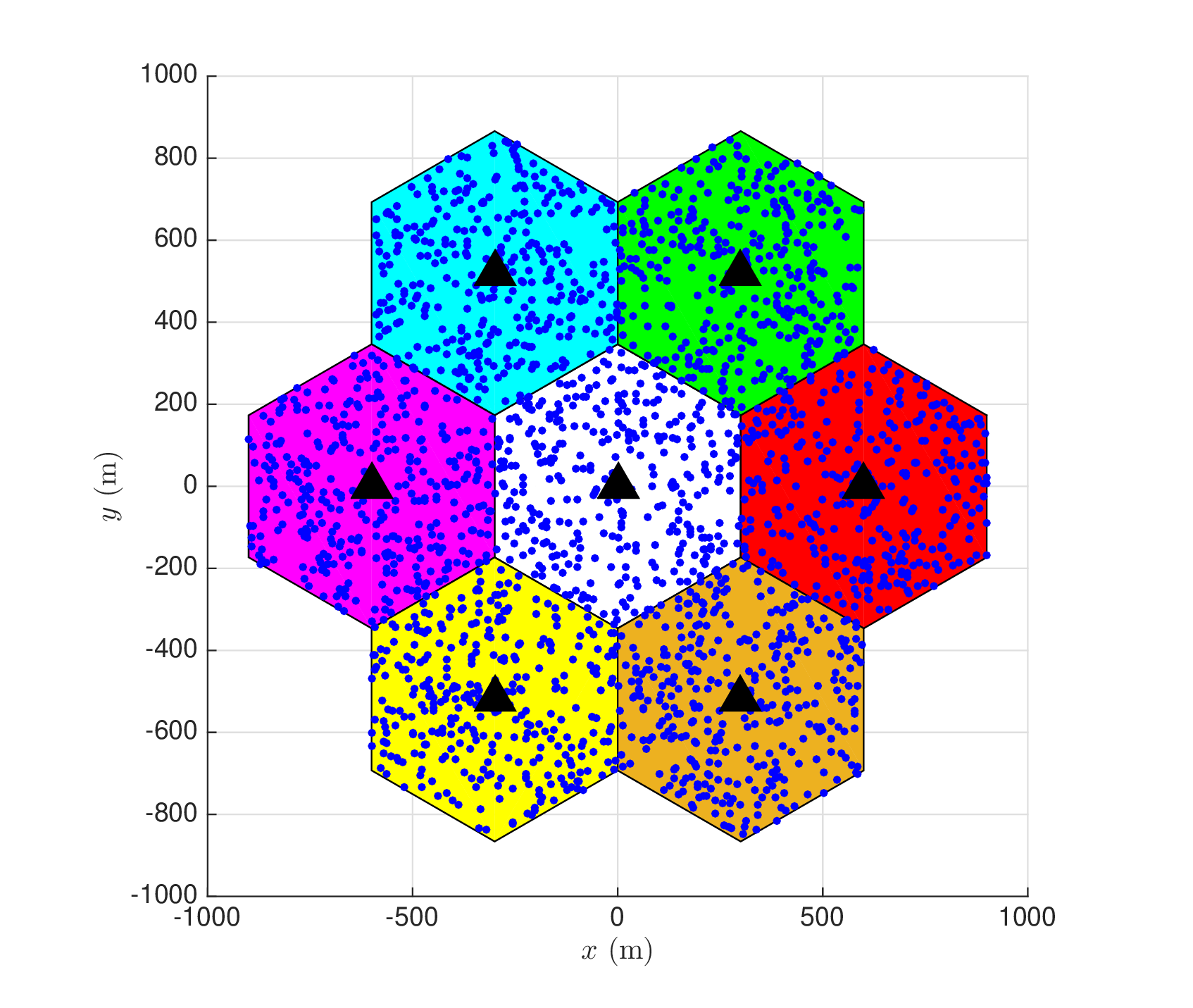}
\par\end{centering}
\caption{Network layout}

\label{fig:layout}
\end{figure}

We consider a region of interest (ROI) under the coverage of a cellular
network where UEs, i.e., wireless devices, are deployed randomly according
to some spatial point process, and BSs cover the entire area. In this
work, for the purpose of developing an analytical model that will
be helpful in illustrating the dependence of the energy consumption
on specific parameters, we assume a homogeneous spatial Poisson point
process (HSPPP) representing the UEs' layout, and a triangular grid
of BSs with hexagonal cells\footnote{The extension of the model to irregular BS deployments is possible,
but it would only complicate our analysis without providing substantial
additional knowledge.}, see Figure~\ref{fig:layout}. Moreover, we assume that the position
of the UEs are fixed, while an adaptation of our solution that takes
into account the impact of mobility on the considered system is left
as future work.

To reliably transmit packets, the BSs and the UEs use power control.
In the case of the UE, this is also controlled by the network operations,
which instruct the UEs on the power they need to use to reach a given
neighbor while sending a packet to it \cite{Yu2009}. This network
assisted power control is made possible by suitable signaling protocols.
In this work, we relate the transmission range to the radio propagation
attenuation, or path loss, through the formulas provided by \cite{METIS_chanmod}\footnote{\cite{METIS_chanmod} provides a widely recognized set of channel
models for different scenarios and communication conditions (I2D,
D2D, V2V, indoor, outdoor, etc..), based on large measurement campaigns
conducted by National telcos, e.g., DoCoMo, in different world cities. }. Specifically, we use the ``UMi-O2O-D2D/V2V'' and ``UMi-O2O-(BS-UE)-LOS''
in Table 7-1 of \cite{METIS_chanmod}. For a given channel gain $g\left(r\right)$,
defined as the inverse of the path loss, signal bandwidth $B$, noise
spectral density $\mathcal{N}_{0}$, noise figure of the receiver
$\eta$ and transmit power $P,$ the channel capacity in bit per second
is $c=B\log_{2}\left(1+Pg\left(r\right)/B\sigma_{c}^{2}\right),$
with $\sigma^{2}=\mathcal{N}_{0}\eta$. Assuming that the packet size
is $K$ bits (including the necessary overhead) and the transmission
takes $T$ seconds, the required transmit power is determined by the
relation $cT=K.$ The energy $\varepsilon$ consumed for the transmission,
given by the product $P\cdot T$, can be expressed, as a function
of the channel gain, as
\begin{equation}
\varepsilon\left(g\left(r\right)\right)=\frac{1}{g\left(r\right)}B\sigma^{2}\left(2^{K/BT}-1\right).\label{eq:energy_consumption}
\end{equation}

\subsection{Traffic model and content request process\label{subsec:Traffic-model}}

In this work, we consider downlink traffic of popular contents. The
traffic demand consists of UE's requesting popular contents generated
by applications at some remote servers, hereafter called ``content
sources'', and the contents are conveyed to the UE's through the
cellular network. Requested contents are grouped in content classes.
Content classes reflect the different types of traffic demand that
may be satisfied by the cellular network with the aid of D2D communications.
In this work, we identify three main features that are important for
characterizing a content class: (i) the content class popularity,
hereafter indicated with the symbol $\phi$, which reflects the percentage
of users that are interested, and hence will sooner or later request,
in each content in a given class, (ii) the request intensity profile,
which determines the lifecycle and the evolution over time of the
frequency of requests for contents in a given class, after they are
generated, and (iii) the delay tolerance, which reflects if, and for
how much time, it is possible to defer the transmission of a content
to a user that has requested it. We identify a content class with
the tuple $\left(\phi,\beta,\kappa,\tau_{c}\right)$. The meaning
of the different symbols is explained below.

\subsubsection{Content class popularity}

As said, the popularity of a content reflects the percentage of users
that will eventually request the content. For implementing this feature,
in this work, we assume that, for each content generated in the class
identified by the tuple $\left(\phi,\beta,\kappa,\tau_{c}\right)$,
the interest of any user for it, i.e., the fact that it will either
request it or ignore it, is the result of a random binary selection:
the user will request the content with probability equal to the popularity
$\phi$ of the content class, which is assumed to be in the range
$[0,1]$. We assume that such binary selections are made independently
across different users and different contents of the same class. It
is straightforward to show that under this assumption, the share of
users that, on average, will request the content during the content
lifecycle (see below), is effectively given by $\phi$. Note that
classes with different request intensity profiles can have the same
popularity $\phi$. Summarizing, \emph{whether or not} a given user
will request a content is determined by the class popularity $\phi$.
\emph{When }the user will request it (if it will do so) is determined
by the request intensity profile $f\left(t;\beta,\kappa\right)$,
described hereafter.

\subsubsection{Request intensity temporal profile}

The users' request intensity temporal profile $f\left(t;\beta,\kappa\right)$
of a class $\left(\phi,\beta,\kappa,\tau_{c}\right)$ describes the
evolution, over time, of the intensity of requests arising from the
users for a given content in the considered class, starting from the
instant the content is generated at the content source. Each content
generated in the class triggers a spatio-temporal offspring of content
requests arising from nodes in the ROI. The profile describes the
dynamics of the number of request per second arising from the UE population
for the content in that class. Although, to the best of our knowledge,
there are no consolidated models for the dynamics of the requests
for contents in the application domains considered in this work, it
is reasonable that, after a content is generated, there is an initial
phase with a low intensity of requests, during which more and more
users acquire awareness, or are notified, of the content existence.
The intensity then increases until a peak is reached, and finally
decays towards zero requests. For the sole purpose of obtaining closed
form results, we use a specific two-parameter template for the request
intensity profile, which is defined as a function of the time $t$
elapsed from the content generation at the content sources, and parameterized
on the scale parameter $\beta$, representing how fast the number
of requests for that content rises and then fades to zero, and the
shape parameter $\kappa$, which defines the behavior of the profile
for small values of $t$, i.e. how aggressively the number of requests
starts to raise. Despite the selection of a specific template, the
approach used in this work can be applicable for any reasonable content
diffusion model. We assume that the request instants by the interested
nodes follow a Poisson temporal distribution with a time-varying average
arrival rate (or temporal density) coinciding with the request intensity
profile $f\left(t;\beta,\kappa\right)$. In general, we observe that
a convenient choice for modeling the request intensity profile is
to impose that $\int_{0}^{\infty}f\left(t;\beta,\kappa\right)dt=1.$
In this way, $f\left(t;\beta,\kappa\right)$ can be also used to represent
the probability density function (PDF) of the instant at which any
user will request the content after it has been generated, conditioned
on the user being interested in it. Along with $f\left(t;\beta,\kappa\right)$,
it is also useful to introduce the corresponding cumulative distribution
function $F_{T}\left(t;\beta,\kappa\right)=\int_{0}^{t}f\left(\tau;\beta,\kappa\right)d\tau$,
which represents the share of interested users that have requested
the content \emph{up to} a given instant $t$ after the content generation.

\subsubsection{Delay tolerance}

The delay tolerance, is the maximum delay, after a UE has requested
a content, within which the content delivery process must guarantee
that the UE receives it. We represent this constraint with the \emph{content
timeout} parameter $\tau_{c}$, which is the fourth element of the
tuple defining a content class. Clearly, for non-delay tolerant traffic,
$\tau_{c}=0$. In general, contents whose fruition is directly related
to a human interaction, e.g., web pages, social media contents, videos,
etc., need to be delivered soon after the content requests as the
user expects to visualize them in a matter of seconds (or less). Other
types of contents, related e.g. to applications running in background
(e.g. automated software updates, context-related information, etc..),
may tolerate delays of minutes or even tens of minutes.

\subsection{Offloading protocol}

We assume that a distributed content delivery management system (CDMS)
is in place at the network operator elements, which coordinates the
actions of BSs and devices in the content delivery process\footnote{This kind of CDMS is a typical assumption in the D2D offloading literature,
see, e.g., \cite{Whitbeck2011,Whitbeck2012,Bruno2014,Pescosolido2018WoWMoM,Pescosolido2018AdHocNetworks}.}. The CDMS implements a simple protocol, described in the following.
Whenever a UE requires a content, a D2D communication is used, in
place of an I2D one, provided that a neighboring UE has already requested
and obtained the content. The D2D communication is between the closest
neighbor of the requesting node that holds a local copy of the content
as the transmitter, and the requesting UE as the receiver.

If, at the time of request, there are no neighbors able to provide
the requested content to the requesting user, we distinguish two cases:
\begin{description}
\item [{A}] In case the content belongs to a \emph{non}-delay tolerant
content class, the cellular infrastructure immediately takes the content
delivery in charge, and transmits it to the requesting device using
a I2D transmission.
\item [{B}] In case the content belongs to a delay tolerant content class,
the cellular network suspends the transmission for an amount of time
defined as ``content timeout'' which starts at the content request
instant. During the content timeout, if a neighbor of the requesting
node receives the content (as a consequence of an independent request
it has issued earlier) it sends the content to the requesting node.
At the expiration of the content timeout, if the requesting node has
not yet received the content, the infrastructure completes the process
by sending it through a I2D communication.
\end{description}
In order to make it available for satisfying future requests from
neighboring nodes, any node that receives the content caches it locally.
In this work, we assume that in the device caches there is always
room for caching newly received contents. In practice, when the physical
storage capacity of a device is reached, newly received contents would
replace older ones based on some criteria. For instance, contents
whose request lifecycle, accordingly to the definition provided in
Subsection~\ref{subsec:Traffic-model}, is complete or near completion,
could be selected as the target of removal\footnote{The effect of a finite caching capacity will be considered in a future
work. }.

\emph{Remark - }In the context of this work, the concept of ``neighboring
devices'' deserves a more detailed description. In most studies two
nodes are considered to be neighbors if they are within a given transmission
range of each other, also known as \emph{contact distance}. This distance,
in turn, is related to the transmit power of the devices, which is
typically associated to physical layer parameters, radio propagation
properties, and/or RF emission limits imposed by the regulators. In
the following, we will show that it is convenient to also link it
to the specific content class. In other words, for us, it is possible
that two nodes are considered neighbors in the context of the delivery
process of a content in some\emph{ specific }class, while at the same
time it is possible that they are \emph{not }neighbors in the context
of the delivery process of a content in another class. This will be
a result of selecting the D2D maximum transmission range as a function
of the content class to pursue the energy consumption minimisation
objective.

\section{Optimal D2D maximum transmission range\label{sec:Problem-formulation}}

We start considering a specific content belonging to class $\left(\phi,\beta,\kappa,\tau_{c}\right)$.
Let $r_{\max}$ be a given maximum D2D transmission range set by the
CDMS for the delivery of contents in this class. We denote with $E\left(r_{\max}\right)$
the average energy spent by the system for delivering the content
(of class $\left(\ensuremath{\phi,\beta},\ensuremath{\kappa},\tau_{c}\right)$)
to an intended user when the maximum D2D transmission range is set
to $r_{\max}$, with $E_{\text{D2D}}\left(r_{\max}\right)$ the component
of $E\left(r_{\max}\right)$ due to the D2D communications, and with
$E_{\text{I2D}}\left(r_{\max}\right)$ the component of $E\left(r_{\max}\right)$
due to the I2D ones. Note that these average values are the ratios
of the sums of the energy spent in the D2D and I2D transmissions,
respectively, to the \emph{overall} number of transmissions, i.e.,
the number of users requesting the content, irrespective of how they
obtain it. The three quantities are related through
\begin{align}
E\left((\ensuremath{\phi,\beta},\ensuremath{\kappa},\tau_{c}),r_{\max}\right)=\, & E_{\text{D2D}}\left((\ensuremath{\phi,\beta},\ensuremath{\kappa},\tau_{c}),r_{\max}\right)\label{eq:average_energy}\\
 & +E_{\text{I2D}}\left((\ensuremath{\phi,\beta},\ensuremath{\kappa},\tau_{c}),r_{\max}\right).\nonumber 
\end{align}

In order to cast our optimisation problem, we define the following
cost function associated to the content delivery process
\begin{align}
C\left(\ensuremath{\phi,\beta},\ensuremath{\kappa},\tau_{c},r_{\max}\right)= & w\cdot E_{\text{D2D}}\left((\ensuremath{\phi,\beta},\ensuremath{\kappa},\tau_{c}),r_{\max}\right)\label{eq:cost_function}\\
 & +\left(1-w\right)E_{\text{I2D}}\left((\ensuremath{\phi,\beta},\ensuremath{\kappa},\tau_{c}),r_{\max}\right),\nonumber 
\end{align}
with $w\in[0,1]$. The value of $r_{\max}$ to select for the delivery
process is
\begin{equation}
\hat{r}_{\max}=\underset{r_{\max}\in[0,\infty)}{\arg\min}\left(C\left(\ensuremath{\phi,\beta},\ensuremath{\kappa},\tau_{c},r_{\max}\right)\right).\label{eq:r_max_opt}
\end{equation}
In \eqref{eq:cost_function}, the weight $w$ represents the availability,
or willingness, of the users to participate in the D2D offloading
process. Low values of $w$ means that the users are available to
spend a considerable amount of energy in the delivery process, and
hence the cost function gives more weight to the I2D energy component.
High values of $w$ make the user privilege their own energy expenditure
minimisation. For simplicity, we assume that a uniform value of $w$
is set for all the users. The selected value may be taught as the
result of incentivizing campaigns targeting the users, which wouldn't
otherwise participate in the D2D offloading process. The analysis
of the responsiveness of the population of users in supporting the
network using D2D communications is outside the scope of this work.
Thus, for us, $w$ is given as an input problem parameter. Note that,
with $w=0.5$, minimising $C\left(\ensuremath{\phi,\beta},\ensuremath{\kappa},\tau_{c},r_{\max}\right)$
is equivalent to minimising the overall energy expenditure \eqref{eq:average_energy},
entailed by I2D and D2D content transmissions.

We proceed by analyzing the different terms in \eqref{eq:cost_function}.
According to the request process defined in Section~\ref{subsec:Traffic-model},
under the assumptions in Section~\ref{subsec:Network-model} on the
network topology, it is easy to check that the location of the UEs
that have \emph{requested} the content \emph{up to }instant $t$ after
the content generation by the content source, is described by a HSPPP
which results from thinning the original HSPPP of the overall nodes'
positions, with a probability value $\phi\cdot F_{T}\left(t;k,\beta\right)$:
first, the overall content popularity $\phi$ (coinciding with the
popularity of its class) restricts the set of UEs that, sooner or
later, will request the content. Then, the CDF $F_{T}\left(t;k,\beta\right)$
determines the fraction of these UEs that have issued the request
up to instant $t$. The resulting HSPPP has spatial density $\rho^{(z,\beta,\kappa)}\left(t\right)$
given by
\begin{equation}
\rho_{z,\beta,\kappa}\left(t\right)=\rho\cdot\phi\cdot F_{T}\left(t;k,\beta\right),\label{eq:device_density_at_run-time}
\end{equation}
 where $\rho$ is the spatial density of the HSPPP of the positions
of all the UEs in the network.

In the following, we first consider non-delay tolerant classes. For
this case, we are able to devise an analytical model that perfectly
matches the simulations results (see Section~\eqref{sec:Performance-evaluation}).
We proceed as follows: first, we compute the probability of offloading
as a function of the content class parameters and $r_{\max}$. Then,
we obtain the probability distribution of the transmission distance
and of the corresponding energy consumption. We then proceed by doing
the same for I2D transmissions. These results will provide all the
terms appearing in Eq.~\eqref{eq:cost_function}. Finally, the optimal
$r_{\max}$ is obtained by minimising \eqref{eq:cost_function} with
respect to $r_{\max}$.

\subsubsection*{Average energy consumption component due to D2D transmissions}

With non-delay tolerant content classes, the content timeout is not
present, and content requests are either fulfilled through D2D immediately
(if there is a neighbor of the requesting user which has already obtained
it) or, still immediately, by a BS using an I2D transmission. The
probability that a content request arriving at instant $t$ is satisfied
through D2D by a device which has already requested (and obtained)
the content, is the probability that, at instant $t$, at least one
user within a range $r$ (from the requesting device) lower than or
equal to the maximum D2D transmission range $r_{\max}$ has requested
the content. This probability coincides with the probability that
the point-to-nearest neighbor distance of the above defined HSPPP
with density $\rho_{z,\beta,\kappa}\left(t\right)$, indicated hereafter
with the random variable $R$, is lower than or equal to $r_{\max}$.
A well known property of HSPPP is that the point-to-nearest neighbor
distance is distributed according to the CDF $F_{R}\left(r;t\right)=1-e^{-\rho_{z,\beta,\kappa}\left(t\right)\pi r^{2}}$
and PDF $p_{R}\left(r;t\right)=2\rho_{z,\beta,\kappa}\left(t_{\text{req}}\right)\pi\cdot re^{-\rho_{z,\beta,\kappa}\left(t\right)\pi r^{2}}$.
Therefore, the probability that the content request can be provided
to the requesting user through a D2D transmission by a neighbor that
has requested the content before $t$ is given by
\begin{align}
\mathbb{P}\left(\text{D2D};\left(\phi,\beta,\kappa\right),r_{\max},t\right) & =\mathbb{P}\left(R\leq r_{\max};t\right)\label{eq:offloading_prob}\\
 & =F_{R}\left(r_{\max};t\right)=1-e^{-\rho_{z,\beta,\kappa}\left(t\right)\pi r_{\max}^{2}}.\nonumber 
\end{align}
The distribution of the transmission distance for D2D transmissions,
is given by the conditional CDF of $R$, with conditioning event $\left\{ R\leq r_{\max}\right\} $.
It is straightforward to show that the corresponding PDF is given
by
\begin{align}
p_{R}\left(r\mid R\leq r_{\max};t\right) & =\frac{2\rho_{z,\beta,\kappa}\left(t\right)\pi\cdot re^{-\rho_{z,\beta,\kappa}\left(t\right)\pi r^{2}}}{1-e^{-\rho_{z,\beta,\kappa}\left(t\right)\pi r_{\max}^{2}}}u_{[0,r_{\max}]}\left(r\right),\label{eq:PDF of D2D tx range}
\end{align}
where the notation $u_{[a,b]}\left(\cdot\right)$ stands for a function
with value 1 in the interval $[a,b]$ (including infinite extremes,
like $[a,\infty)$ and open borders, like $(a,b]$) and zero outside.

Note that the intensity of requests, the probability of offloading,
and the distribution of the D2D transmission distance vary with time
as a function of the density of the devices that have the content
in their caches, which progressively increases. Therefore, to obtain
the component of the average energy consumption associated to D2D
transmissions, it is necessary to integrate, over the time domain,
the product of the intensity of the requests times the time-varying
probability of offloading, times the energy consumption averaged over
the distribution of the transmission distance. Using the model defined
in \eqref{eq:energy_consumption}, the energy consumption component
due to D2D communications can be obtained as 
\begin{align}
E_{\text{D2D}}\left(r_{\max}\right)= & \int_{0}^{\infty}f\left(t;k,\beta\right)\mathbb{P}\left(\text{D2D};\left(\phi,\beta,\kappa\right),r_{\max},t\right)\label{eq:D2D_av_energy}\\
 & \cdot\int_{0}^{r_{\max}}p_{R}\left(r\mid R\leq r_{\max};t\right)\varepsilon\left(g_{\text{D2D}}\left(r\right)\right)drdt,\nonumber 
\end{align}
which, using \eqref{eq:PDF of D2D tx range} and \eqref{eq:offloading_prob}
simplifies to
\begin{align}
E_{\text{D2D}}\left(r_{\max}\right)= & \int_{0}^{\infty}f\left(t;k,\beta\right)\\
 & \cdot\int_{0}^{r_{\max}}\hspace{-4mm}2\rho_{z,\beta,\kappa}\left(t\right)\pi\cdot re^{-\rho_{z,\beta,\kappa}\left(t\right)\pi r^{2}}\varepsilon\left(g_{\text{D2D}}\left(r\right)\right)drdt,\nonumber 
\end{align}
where $g_{\text{D2D}}\left(r\right)$ is the channel gain for D2D
channels.

\subsubsection*{Average energy consumption component due to I2D transmissions}

Under the assumption that the spatial point process of the nodes layout
is homogeneous, the position of the receiving device of any I2D transmission
is uniformly distributed over the cell surface. For the purposes of
this work, we assume hexagonal cells. Under this assumption, it can
be showed that the distance of a device from the BS, located at the
center of the cell, is distributed according to the following PDF:
\begin{align}
p_{R_{\text{I2D}}}\left(r\right)= & \frac{r}{r_{\text{out}}^{2}\sin\frac{\pi}{3}}\left(\frac{\pi}{6}u_{[0,r_{\text{out}}\sin\frac{\pi}{3}]}\left(r\right)\right.\label{eq:I2D_TX_distance_PDF}\\
 & \left.+\left(\arcsin\left(\frac{r_{\text{out}}}{r}\sin\frac{\pi}{3}\right)-\frac{\pi}{3}\right)u_{(r_{\text{out}}\sin\frac{\pi}{3},r_{\text{out}}]}\left(r\right)\right),\nonumber 
\end{align}
where $r_{\text{out}}$ is the maximum cell radius, i.e. the distance
of the cell center from any hexagon vertex.

From the PDF of the I2D transmission range, it is straightforward
to compute the I2D component of the energy consumption as
\begin{align}
E_{\text{I2D}}\left(r_{\max}\right)= & \int_{0}^{\infty}f\left(t;k,\beta\right)\left(1-\mathbb{P}\left(\text{D2D};\left(\phi,\beta,\kappa\right),r_{\max},t\right)\right)dt\nonumber \\
 & \cdot\int_{0}^{r_{\text{out}}}p_{R_{\text{I2D}}}\left(r\right)E\left(g_{\text{I2D}}\left(r\right)\right)dr,\label{eq:I2D energy component}
\end{align}
where $g_{\text{I2D}}\left(r\right)$ is the channel gain for I2D
channels.

Equations \eqref{eq:D2D_av_energy} and \eqref{eq:I2D energy component}
allow to compute the cost function \eqref{eq:cost_function}, and,
in turn, the optimal $r_{\max}$ in \eqref{eq:r_max_opt}. The computation
turns out to be very simple. In fact, it can be shown that cost function
defined in \eqref{eq:cost_function} has a global minimum, as it is
the sum of two terms with opposite behavior as a function of $r_{\max}$.
The first term (related to the D2D energy consumption) keeps increasing
as $r_{\max}$ increases, as both the probability of offloading and
the average energy per D2D transmission, increase with $r_{\max}$.
On the contrary, the second term in \eqref{eq:cost_function}, associated
to I2D transmissions, keeps decreasing with $r_{\max}$, as the probability
$1-\mathbb{P}\left(\text{D2D};\left(\phi,\beta,\kappa\right),r_{\max},t\right)$
appearing in \eqref{eq:I2D energy component} decreases, while the
term $\int_{0}^{r_{\text{out}}}p_{R_{\text{I2D}}}\left(r\right)E\left(g_{\text{I2D}}\left(r\right)\right)dr$
is not affected by it. Therefore, any basic minimum search method
for the global minimum of a function of a scalar variable can be used
to find the optimum $r_{max}$ in a computationally light way.

The devised analytical model perfectly covers the case of non-delay
tolerant traffic. With delay tolerant traffic, the content delivery
from the BS is deferred of $\tau_{c}$ seconds (the content timeout)
with respect to an incoming request, unless a neighbor of the requesting
user is found with the content in cache in the meantime. An analytical
model for the I2D and D2D components of the energy consumption is
more difficult to obtain with respect to the case for non-delay tolerant
traffic. In fact, the computation of the probability of offloading
requires the knowledge of the runt-time density of devices with the
content in cache. For content timeouts of duration comparable with
the timescale of the request intensity profile, the run-time density
of devices that have \emph{received} the content at some instant $t$
may diverge in a non-negligible way from the density $\rho_{z,\beta,\kappa}\left(t\right)$
in \eqref{eq:device_density_at_run-time} of the devices that have
requested it up to instant $t$. Moreover, the evolution of the presence
of contents in the devices' caches presents strong spatial correlation,
which makes the problem non tractable with the tools of HSPPPs. For
the purpose of this work, which is to show how beneficial a selective
setting of the maximum D2D transmission range can be in enhancing
the energy saving capabilities of D2D offloading techniques, we cover
the non-delay tolerant case by means of simulations, as showed in
the next section.\vspace{-2mm}

\section{Performance evaluation\label{sec:Performance-evaluation}\vspace{-1mm}
}

\begin{figure}[t]
\centering{}\includegraphics[width=0.9\columnwidth]{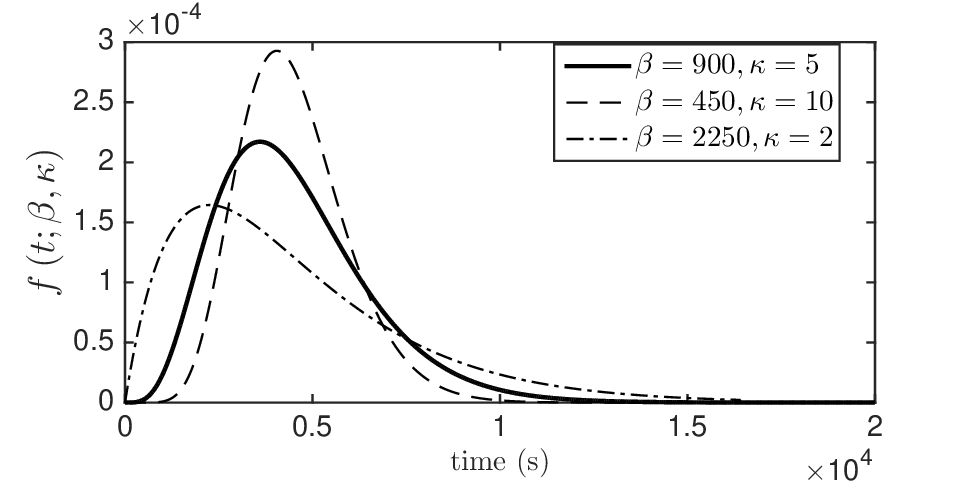}\vspace{-2mm}
\caption{Request intensity profile for different parameters $\beta$ and $\kappa$.}
\label{fig:req_intensity_profile}\vspace{-6mm}
\end{figure}

We validate our model and illustrate the system performance using
simulation results obtained with a custom Matlab simulator. The ROI
is a surface covered by 7 hexagonal cells of inner radius 300m (see
Figure~\ref{fig:layout}). The device density is $1.1\cdot10^{-3}$
devices/m$^{2}.$ Each result has been obtained by averaging over
100 independent realizations of the devices layout. To avoid border
effects on the statistics, although the simulations cover the activity
of all the nodes in the ROI, we computed the statistics using the
data related to the central cell only (which can receive contents
from neighbors in all directions). To generate the content request
process, we used, as a template for the request intensity profile
$f\left(t;\beta,\kappa\right)$, the PDF of the Gamma distribution
with rate parameter $\beta$ and shape parameter $\kappa$:
\[
f\left(t;\beta,\kappa\right)\triangleq\frac{\beta^{\kappa}}{\Gamma\left(\kappa\right)}t^{\kappa-1}e^{-\beta t},
\]
where $\Gamma\left(\kappa\right)=\int_{0}^{\infty}t^{\kappa-1}e^{-t}dt$
is the Gamma function. For practical reasons, we truncated the profile
(and renormalized to obtain a proper PDF) at $t=20000$ seconds (around
5 hours and a half). We set the shape parameter as $\kappa=5$, and
the scale parameter $\beta=900$. In this way, we obtain the profile
shown in Figure~\ref{fig:req_intensity_profile} with the tick line,
along with different profiles obtained with different setting of the
parameters. The selected profile peaks at 1 hour after the content
generation, and in less then 3 hours (174 mins), 99\% of the interested
users (on average) have requested the content.

To compute the energy consumption of packet transmissions, we have
assumed a bandwidth $B=$1 MHz and a packet transmission duration
of $T=1$ seconds. In this work, we assume that the radio resource
management layer of the cellular network allocates orthogonal radio
resources, e.g., physical resource block (PRBs) of a time-frequency
grid, to concurring transmissions, to avoid interference. The term
$\sigma^{2}$ in \ref{eq:energy_consumption} is computed as $\sigma^{2}=\mathcal{N}_{0}\eta$,
with noise spectral density $\mathcal{N}_{0}=-174$~dBm/Hz and noise
figure $\eta=11$~dBm/Hz.\vspace{-2mm}

\subsection{D2D-aided content delivery for a specific content class}

\begin{figure}[t]
\centering{}\includegraphics[width=0.9\columnwidth]{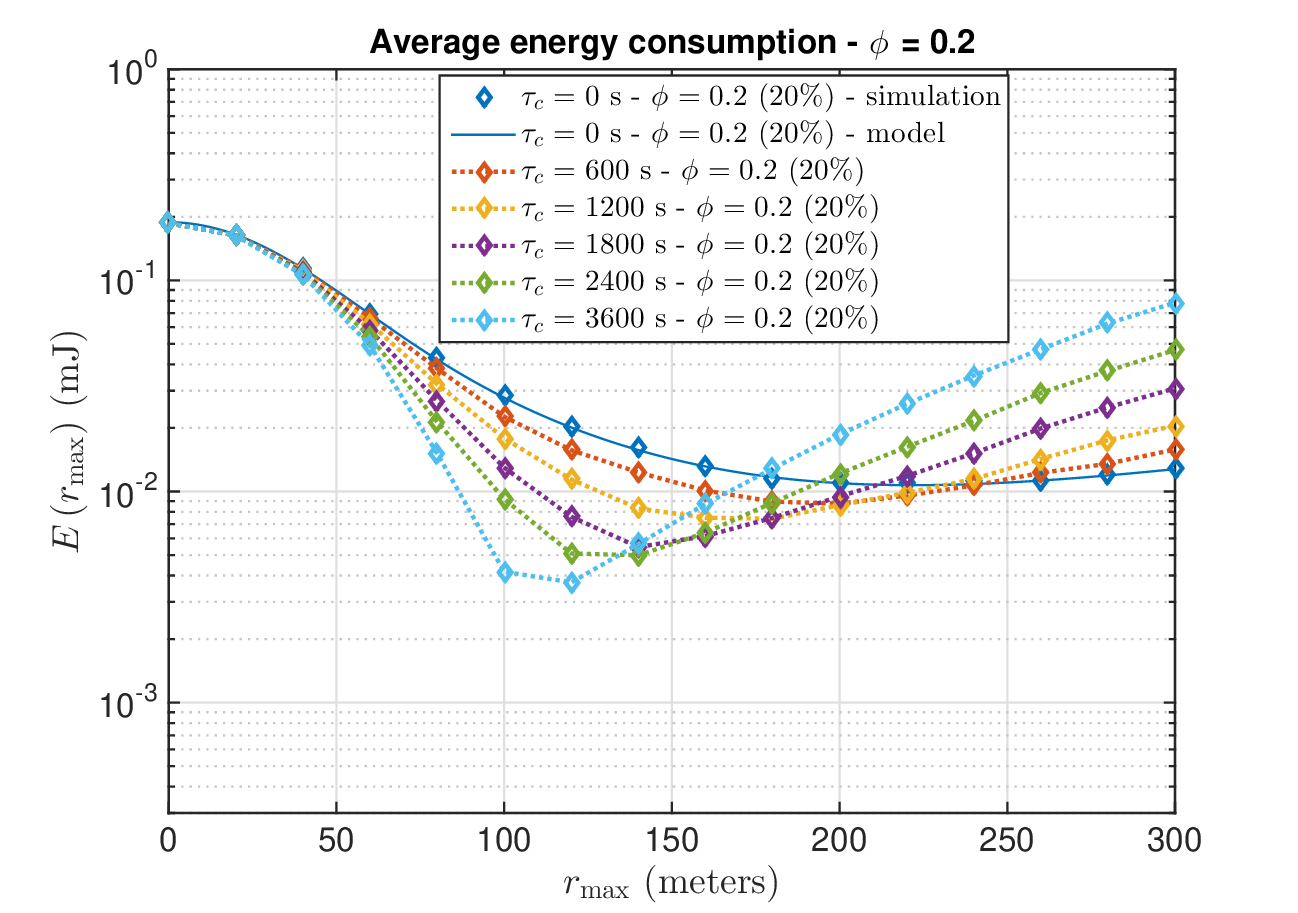}\vspace{-4mm}
\caption{Energy consumption per content delivery with $w=0.5$, $\mathbf{\phi=0.2},\,\beta=900,\,\kappa=5$.}
\label{fig:F1-1}\vspace{-5mm}
\end{figure}

\begin{figure}[t]
\begin{centering}
\includegraphics[width=0.9\columnwidth]{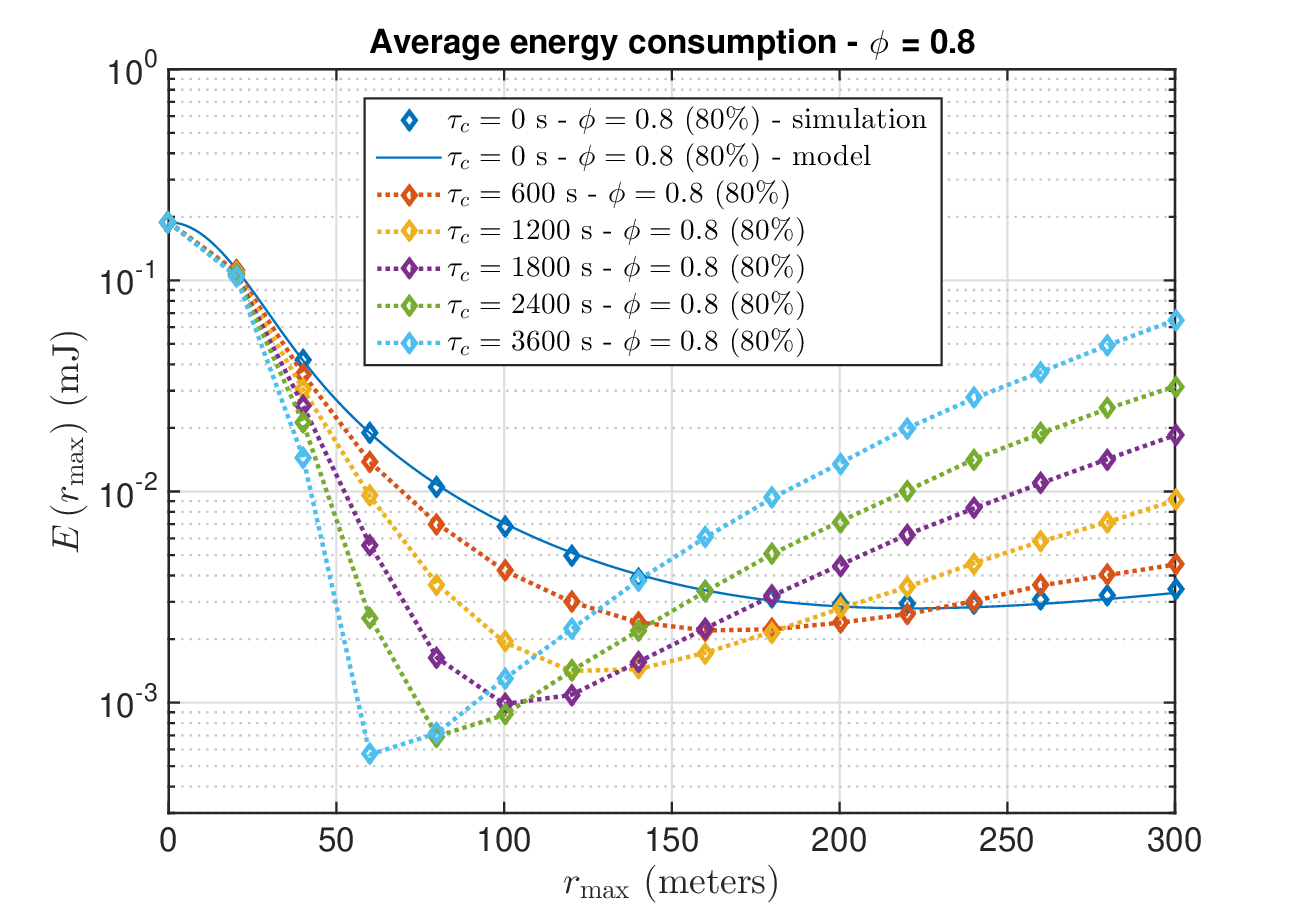}xxx\vspace{-3mm}
\caption{Energy consumption per content delivery with $w=0.5$, \textbf{$\mathbf{\phi=0.8}$},
$\beta=900,\,\kappa=5.$}
\label{fig:F1-5}
\par\end{centering}
\centering{}\vspace{-3mm}
\end{figure}

Figures~\ref{fig:F1-1} and \ref{fig:F1-5} show the average energy
consumption per content delivery for different content classes, as
a function of the selected $r_{\max}$. Figure~\ref{fig:F1-1} refers
to classes with popularity parameter $\phi=0.2$, i.e., 20\% popularity,
and delay tolerance ranging from 0 to 1 hour. Figure~\ref{fig:F1-5}
refers to classes with 8, i.e., 80\% popularity, and delay tolerances
the same range. All the classes have in common the same request intensity
profile, with $\beta=900,\,\kappa=5$. The curves plotted for the
classes with no delay tolerance ($\tau_{c}=0$) report both the values
predicted by the proposed analytical model (solid lines) and the results
of simulations (markers). The other curves in both figures report
the sole results of the simulations. It can be seen that the analytical
model perfectly matches with the simulation results. The descending
and then ascending behavior of this curves is motivated by two effects:
first, with very low values of $r_{\max}$, D2D delivery opportunities
are very rare, and the generally more energy expensive I2D transmissions
prevail. Then, increasing $r_{\max}$ entails an increase of D2D opportunities,
while the required power is still much lower than what would be required
(on average) by I2D transmissions. The resulting effect is an initial
decrease of energy consumption associated to the increase of $r_{\max}$.
However, keeping increasing $r_{\max}$ starts to make the average
D2D energy consumption significant, as there is an increase of both
the number of D2D transmission opportunities and, most importantly,
of their required (on average) transmit power. This causes the curve
to increase.

Setting $w=0.5$ in \eqref{eq:I2D energy component} would allow to
minimise the overall energy (D2D+I2D), i.e., with reference to the
figures, would select the maximum transmission range $r_{max}$, on
a per-class basis, as the one that entails the minimum energy consumption.
It can be seen that a different selection of $r_{\max}$ could entail
a significant increase of the overall energy expenditure. As a result,
any common choice of $r_{\max}$ across all the content-classes would
be suboptimal for handling contents for most of them.\vspace{-2mm}

\subsection{D2D-aided content delivery with heterogeneous traffic\vspace{-1mm}
}

In realistic scenarios, contents in the interest of multiple users
typically belong to content classes differing in terms of popularity,
request intensity profile, tolerable delivery delay, and content generation
rate. In the following, we show that, choosing the optimal maximum
D2D-range selectively, as a function of the parameters of the class
each content belongs to, allows to achieve considerable energy savings
with respect to a uniform maximum D2D-range selection.

\begin{figure}[t]
\centering{}\includegraphics[width=0.9\columnwidth]{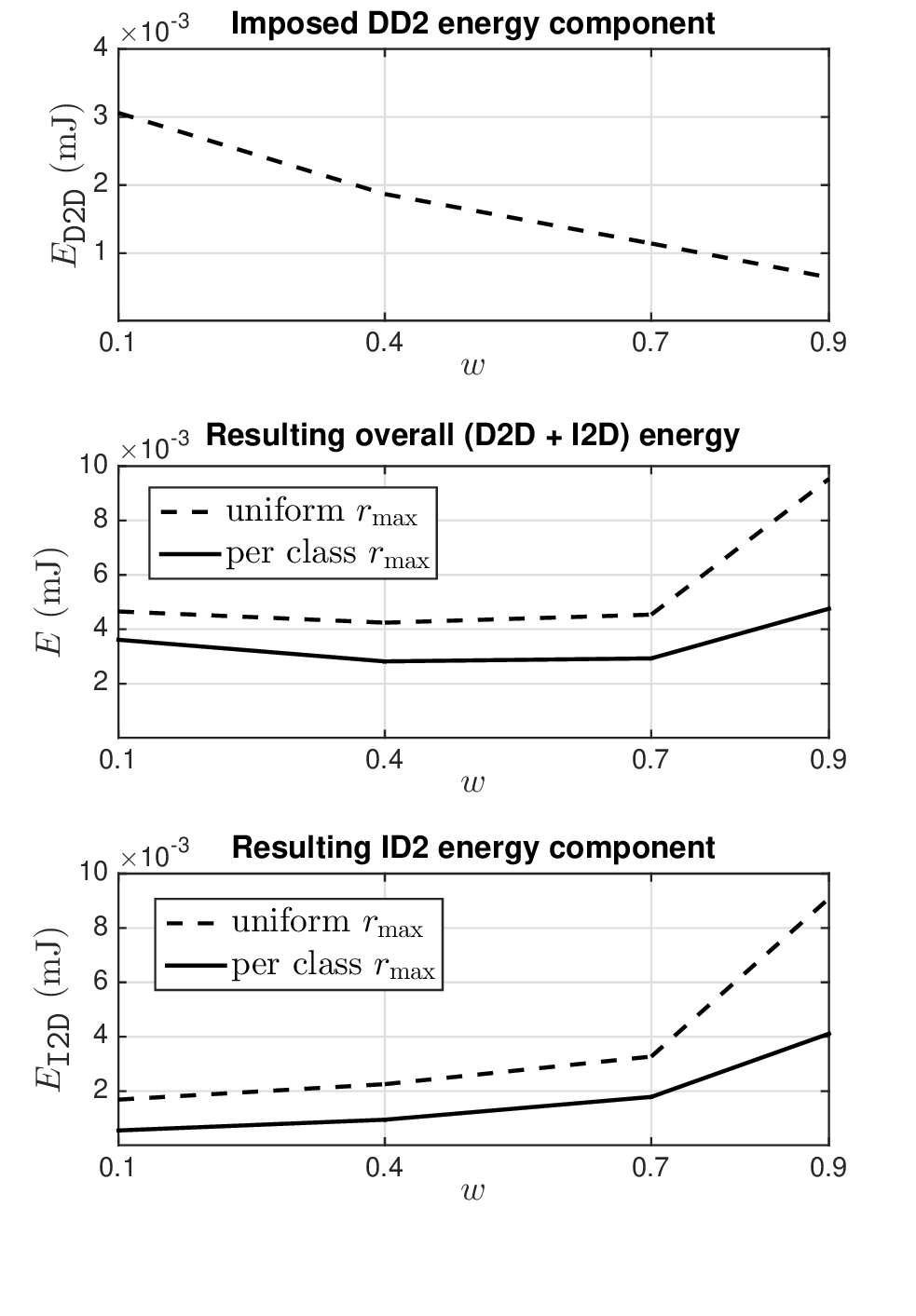}\vspace{-12mm}
\caption{Infrastructure energy consumption with different constraints on the
D2D energy consumption component.}
\label{fig:composite_traffic_1}\vspace{-5mm}
\end{figure}

\begin{figure}[t]
\centering{}\includegraphics[width=0.9\columnwidth]{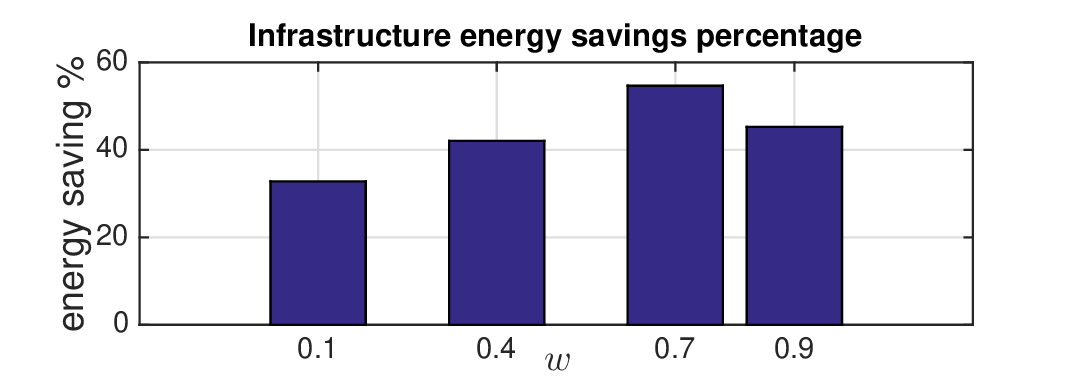}\vspace{-3mm}
\caption{Infrastructure energy savings percentage of the proposed selective
$r_{\max}$ setting strategy with respect to the uniform selection strategy.}
\vspace{-5mm}
\label{fig:composite_traffic_2}
\end{figure}

The results presented in Figures~\ref{fig:composite_traffic_1} and
\ref{fig:composite_traffic_2} were obtained assuming that the overall
traffic load is composed of the content classes determined by any
combination of the parameters listed in Table~\ref{tab:1}, while
we kept common values of the request intensity profile parameters
$\beta=900$ and $\kappa=5$, which entail a request peak one hour
after the generation of each content.\vspace{-2mm}
\begin{table}[h]
\centering{}\caption{Example of traffic load composition}
\vspace{-2mm}
\label{tab:1}%
\begin{tabular}{|c|c|c|}
\hline 
parameter & symbol & values\tabularnewline
\hline 
\hline 
content timeout & $\tau_{c}$ & \{0, 10, 30, 60\} min\tabularnewline
\hline 
popularity & $\phi$ & \{0.2, 0.4, 0.6, 0.8\}\tabularnewline
\hline 
\end{tabular}
\end{table}
 
The number of classes was 16, and we assumed that they shared the
same percentage of the overall traffic load, i.e., 6.25\%, in terms
of number of generated contents (while the number of \emph{delivered}
contents, and hence number of required transmissions, is clearly different
due to the different popularity across the content classes). The top
plot of Figure~\ref{fig:composite_traffic_1} shows the value imposed
as a constraint for the D2D component of the energy consumption. The
values of the D2D energy constraints we considered, correspond to
those obtained with $w\in\left\{ 0.1,\,0.4,\,0.7,\,0.9\right\} $.
The second and third subplots show the corresponding overall energy
consumption (middle subplot) and infrastructure energy consumption
component (bottom subplot), obtained with the proposed solution for
$r_{\max}$ (solid lines), and with the best \emph{common} $r_{\max}$
selection, which is suboptimal for any specific class (dashed lines)\footnote{In order to obtain the performance of the benchmark system, we had
to off-line compute the optimal common choice for $r_{\max}$ (which
is suboptimal for any specific class), defining a cost function similar
to \eqref{eq:cost_function} but applied to the I2D and D2D contribution
related to the aggregate traffic.}. 

Figure~\ref{fig:composite_traffic_2} reports the same results, but
highlighting the percentage reduction of the I2D energy expenditure
compared to the benchmark. It can be seen that for every value of
$w$ the energy saving exceed 30\%, with a peak 55\%.\vspace{-2mm}

\section{Conclusion\label{sec:Conclusion}\vspace{-1mm}
}

The major contribution of this work is the recognition that the transmission
range of D2D communications (or equivalently, their transmit power)
is a crucial parameter in the effort to extract as much as possible
energy savings from the use of D2D communication in support of 5G
and next generation networks. A suboptimal choice of $r_{\max}$,
oblivious to the traffic type, i.e., to the content classes parameters,
may result in energy consumption order of magnitude larger than the
minimum, in a specific content class. As an intermediate theoretical
result, we have also presented an analytical model which comes in
handy to off-line compute the best selection for non-delay tolerant
traffic, and which we plan to extend to the more general case in a
future work. Key steps forward in this study are also the evaluation
of the effect of mobility, including mixed pedestrian and vehicular
scenarios, and the investigation of the achievable gains in terms
of spectrum usage.\vspace{-2.1mm}

\begin{acks}
This work was partially funded by the EC H2020 Research Infrastructure "SLICE DS" Project, under grant no. 951850.\vspace{-2.1mm}
\end{acks}

\bibliographystyle{ACM-Reference-Format}

\end{document}